\documentclass[12pt]{article}
\usepackage{amsmath,amssymb}
\usepackage{graphicx}
\usepackage{cite}
\usepackage{fancyhdr}

\usepackage[affil-it]{authblk}
\usepackage[left=2cm,right=2cm,top=3cm,bottom=3cm,a4paper]{geometry}

\begin{document}

\allowdisplaybreaks
\setcounter{footnote}{0}
\setcounter{figure}{0}
\setcounter{table}{0}

\title{\bf \large 
Muon $g-2$ in MSSM Gauge Mediation Revisited}
\author[1,2]{{\normalsize  Tsutomu T. Yanagida}}
\author[3]{{\normalsize Norimi Yokozaki}}

\affil[1]{\small 
Kavli Institute for the Physics and Mathematics of the Universe (WPI),
 
University of Tokyo, Kashiwa 277--8583, Japan}

\affil[2]{\small Hamamatsu Professor}

\affil[3]{\small 
Department of Physics, Tohoku University,  

Sendai, Miyagi 980-8578, Japan}

\date{}

\maketitle

\thispagestyle{fancy}
\rhead{ IPMU17-0053 \\ TU-1044 }
\cfoot{\thepage}
\renewcommand{\headrulewidth}{0pt}

\begin{abstract}
\noindent
The Higgs boson of 125 GeV requires large stop masses, 
leading to the large $\mu$-parameter in most cases of gauge mediation. 
On the other hand, the explanation for the muon $g-2$ anomaly needs small slepton and neutralino/chargino masses.
Such disparity in masses may be obtained from a mass splitting of colored and non-colored messenger fields. 
However, even if the required small slepton and neutralino/chargino masses are realized, 
all parameter regions consistent with the muon $g-2$ are excluded 
by the recent updated ATLAS result on the wino search 
in the case that the messenger fields are in ${\bf 5}+\bar {\bf 5}$ representations of $SU(5)$. It is also revealed that the messenger fields in ${\bf 10} + \overline{\bf 10}$ or ${\bf 24}$ representation can not explain the muon $g-2$ anomaly.
We show, giving a simple example model, that the above confliction is solved if there is an additional contribution to the Higgs soft mass which makes the $\mu$-parameter small. 
We also show that the required Higgs $B$-term for the electroweak symmetry breaking is consistently generated by radiative corrections from gaugino loops.

%
%
%
\end{abstract}

\clearpage

\section{Introduction}

The anomalous magnetic moment of the muon (muon $g-2$) was precisely measured by the Brookhaven E821 experiment~\cite{Bennett:2006fi,Roberts:2010cj}. The measured value deviates from the standard model (SM) prediction 
as~\cite{Hagiwara:2011af}
\begin{eqnarray}
(a_\mu)_{\rm exp} - (a_{\mu})_{\rm SM} = (26.1 \pm 8.0) \times 10^{-10}, \label{eq:gm2}
\end{eqnarray}
which is at more than 3$\sigma$ level. Also, Davier et al. has reported a deviation of 3.6$\sigma$ level~\cite{Davier:2010nc,Davier:2016iru}. 
This anomaly might be a signal of the physics beyond the SM. 
One of the attractive explanations for the muon $g-2$ anomaly is given by supersymmetry (SUSY). In the minimal SUSY extension of the SM (MSSM), the anomaly of the muon $g-2$ is explained with light sleptons, neutralino and/or chargino of $\mathcal{O}$(100)\,GeV~\cite{Lopez:1993vi,Chattopadhyay:1995ae,Moroi:1995yh}.

Such a low-scale SUSY is, in general, very severely constrained by flavor changing neutral-current processes because of the lightness of the sleptons. 
This fact motivates us to consider SUSY breaking mediation models where soft SUSY breaking masses arise in a flavor independent way. 
In gauge mediated SUSY breaking (GMSB) scenarios~\cite{gmsb1,gmsb2,gmsb3} (for early attempts~\cite{gmsbold1,gmsbold2,gmsbold3,gmsbold4,gmsbold5}),  
the SUSY breaking of the hidden sector is mediated to the visible sector via gauge interactions,
 and the generated soft masses are flavor independent; therefore, GMSB models are attractive candidates for the low-scale SUSY.

In most GMSB models, the Higgs boson mass of 125 GeV requires heavy stops of masses $\sim$10\,TeV~\cite{Okada:1990vk,Ellis:1990nz,Haber:1990aw,Okada:1990gg,Ellis:1991zd}, 
due to the smallness of trilinear couplings of the stops.
With the heavy stops, the higgsino mass parameter, $\mu$, should be taken as large as several TeV for generating the correct electroweak symmetry breaking (EWSB). 
In these cases, the chargino contribution to the muon $g-2$ is suppressed and the explanation of the muon $g-2$ anomaly
requires the quite light sleptons, bino and wino.\footnote{
The heavier wino makes the slepton masses larger via radiative corrections.
} 
The mass splitting of the colored (squarks and a gluino) and non-colored SUSY particles (sleptons, neutralinos and charginos) may be realized from a mass splitting of colored and non-colored messenger fields~\cite{Sato:2012bf,Ibe:2012qu,Bhattacharyya:2013xba,Bhattacharyya:2013xma}.  However, the lightest neturalino/chargino is almost purely wino-like for the large $\mu$-parameter, and its mass is predicted below 400 GeV. This wino has been excluded by the recent updated ATLAS result on the wino search~\cite{ATLAS_wino}: the wino should be heavier than 430\,GeV. This confliction can be avoided in the case of an adjoint messenger transforming as {\bf 24} in $SU(5)$ gauge group. We will show that this case is, however, also excluded since the stau is too light and long-lived (see \cite{lhcstau, lhcstau2} for the LHC bounds). So far, in the cases that the messenger fields are in ${\bf 5}+\bar{\bf 5}$, ${\bf 10} + \overline{\bf 10}$ or ${\bf 24}$ representation of $SU(5)$, the muon $g-2$ anomaly can not be explained without an extension.

The above problem originates from the large $\mu$-parameter. However, the $\mu$-parameter can be taken small as several hundred GeV if  we have  an additional source to the Higgs soft masses~\cite{Ibe:2012qu}. In this letter, we show that the muon $g-2$ anomaly can be easily explained  in a simple model where we have additional Higgs soft masses.
%

\section{Muon $g-2$ in gauge mediation}

\subsection{Preliminaries}

In this subsection, we show that it is impossible to explain the muon $g-2$ consistently with the updated result of the LHC wino search  without an additional Higgs soft mass. 
The difficulty originates from the fact that $\mu$-parameter must be large due to the heavy stops. The lightest neutralino/chargino is purely wino-like. This is because the tree-level mass splitting between charged and neutral winos is highly suppressed for large $\mu$ and large $\tan\beta$~\cite{Gherghetta:1999sw} and the mass splitting dominantly come 
from radiative corrections~\cite{Wells:2004di,Cheng:1998hc,Feng:1999fu,Ibe:2012sx}. 
Here, $\tan\beta$ is the ratio of the vacuum expectation value of the up-type Higgs to that of the down-type Higgs, $\tan\beta=\left<H_u\right>/\left<H_d\right>$. 
In the following, we consider GMSB models with ${\bf 5}+\bar {\bf 5}$, ${\bf 10}+\overline{\bf 10}$ and ${\bf 24}$ messenger fields.

First, we consider the gauge mediation model with messenger fields which form a pair of  ${\bf 5}$ and $\bar {\bf 5}$ representation of $SU(5)$ GUT gauge group. The superpotential is 
\begin{eqnarray}
W = (M_D + k_D Z) \Psi_D \Psi_{\bar D} +(M_L + k_L Z) \Psi_L \Psi_{\bar L},
\end{eqnarray}
where $\Psi_{\bar D}$ and $\Psi_L$ are transformed as ($\bar{\bf 3}$,{\bf 1},1/3) and ({\bf 1},{\bf 2},$-1/2$) under the SM gauge group, $SU(3)_C \times SU(2)_L \times U(1)_Y$; $Z$ is a SUSY breaking field with $\left<Z\right>=0$ and $\left<F_Z\right> \neq 0$. 
After integrating out the messenger fields, the soft SUSY breaking masses are generated as
\begin{eqnarray}
M_1 \simeq \frac{g_1^2}{16\pi^2} \left(\frac{3}{5} \Lambda_L + \frac{2}{5} \Lambda_D \right), \ \ 
M_2 \simeq \frac{g_2^2}{16\pi^2} \Lambda_L,  \ \ 
M_3 \simeq \frac{g_3^2}{16\pi^2} \Lambda_D,  \label{eq:gauginos}
\end{eqnarray}
\begin{eqnarray}
m_{i}^2 \simeq \frac{2}{(16\pi^2)^2} \left[ {C_2(r_3)} g_3^4 \Lambda_D^2  + C_2(r_2) g_2^4 \Lambda_L^2 + \frac{3}{5}Q_Y^2 g_1^4 \tilde \Lambda_5^2 \right],
\end{eqnarray}
where $M_1$, $M_2$ and $M_3$ are the bino, wino and gluino mass, respectively, and $m_i^2$ is a scalar mass squared of a field $i$; $g_1$, $g_2$ and $g_3$ are the gauge coupling constants of $U(1)_Y$, $SU(2)_L$ and $SU(3)_C$; 
$C_2(r_a)$ is the quadratic Casimir invariant of representation $r_a$; $Q_Y$ is a hypercharge. 
Here,  
$\Lambda_L= k_L \left<F_Z\right>/M_L$, $\Lambda_D= k_D \left<F_Z\right>/M_D$ and $\tilde \Lambda_5^2 = (3/5) \Lambda_L^2 + (2/5) \Lambda_D^2$.

For $\Lambda_D \gg \Lambda_L$, the sleptons become much lighter than the squarks, which is needed for the muon $g-2$ explanation. As a result, the wino mass is smaller than the bino mass in the regions consistent with the muon $g-2$ experiment (see Eq.\,(\ref{eq:gauginos})).

The Higgs boson mass of 125 GeV requires heavy stops, leading to the large $\mu$-parameter for the correct electroweak symmetry breaking as
\begin{eqnarray}
\mu^2 \approx -m_{H_u}^2 \sim \frac{3 y_t^2}{4\pi^2} m_{\tilde t}^2 \ln \frac{M_{D}}{m_{\tilde t}} \sim ((3\,{\mathchar`-}\,4) {\rm TeV})^2,
\end{eqnarray}
where $m_{\tilde t}$ is a stop mass of $\sim 10$\,TeV.
With the large $\mu$-parameter, the chargino contribution to the muon $g-2$ is suppressed and 
the bino contribution is dominant. The bino contribution is estimated as~\cite{Endo:2013lva} 
\begin{eqnarray}
(\delta a_\mu)_{\rm bino} \simeq \frac{3}{5}\frac{g_1^2}{16\pi^2} \frac{m_\mu^2 M_1 \mu}{m_{L_2}^2 m_{\bar E_2}^2}
\tan\beta \cdot  f_N \left(
\frac{m_{L_2}^2}{M_1^2},
\frac{m_{\bar E_2}^2}{M_1^2}
 \right),
\end{eqnarray}
which is enhanced for the large $\mu\tan\beta$. Here, $m_{L_2}$ ($m_{\bar E_2}$) is a left-handed (right-handed) smuon mass and $f_N$ is a loop function. 
In order to explain the muon $g-2$ anomaly, both of the left-handed and right-handed sleptons need to be lighter than $\sim$\ 500 GeV for $\mu\tan\beta \approx 120$\,TeV. The light left-handed sleptons can be obtained by the small $\Lambda_L$. However, it leads to the light wino, which is easily excluded by the LHC wino search using a disappearing track~\cite{ATLAS_wino}: the wino should be heavier than 430\,GeV. Another difficulty is that $\Psi_{\bar  D}$ has a hypercharge of 1/3, which prevents the sleptons from being sufficiently light. 

Moreover, the large $\mu \tan\beta$ is problematic, since it in turn generates the large left-right mixing of the staus: 
when the left-right mixing is large, the stau-Higgs potential has a charge breaking minimum which can be deeper 
than the EWSB minimum. Requiring that the life-time of the EWSB minimum be longer than the age of the universe, we have an upper-bound on $\mu\tan\beta$ as~\cite{Hisano:2010re}
\begin{eqnarray}
\mu \tan\beta &<& 213.5 \sqrt{m_{L_3} m_{\bar E_3}} -17.0 (m_{L_3} + m_{\bar E_3})  \nonumber \\
&+& 4.52 \times 10^{-2}\, {\rm GeV}^{-1} (m_{L_3} -m_{\bar E_3})^2 - 1.30 \times 10^4\,{\rm GeV}, \label{eq:chbreaking}
\end{eqnarray}
where $m_{L_3}$ ($m_{\bar E_3}$) is a mass of a left-handed (right-handed) stau. Since $m_{L_3}$ and $m_{\bar E_3}$ need to be smaller than $\sim$500\,GeV, this leads to the tension between the stability of 
the charge conserving EWSB minimum and the solution to the muon $g-2$ anomaly.

\begin{figure}[!t]
\begin{center}
\includegraphics{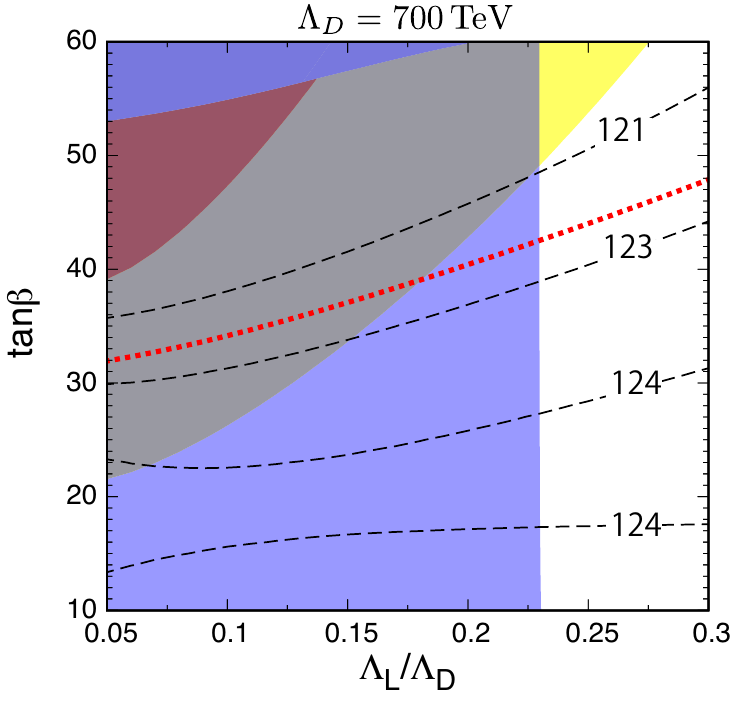}
\includegraphics{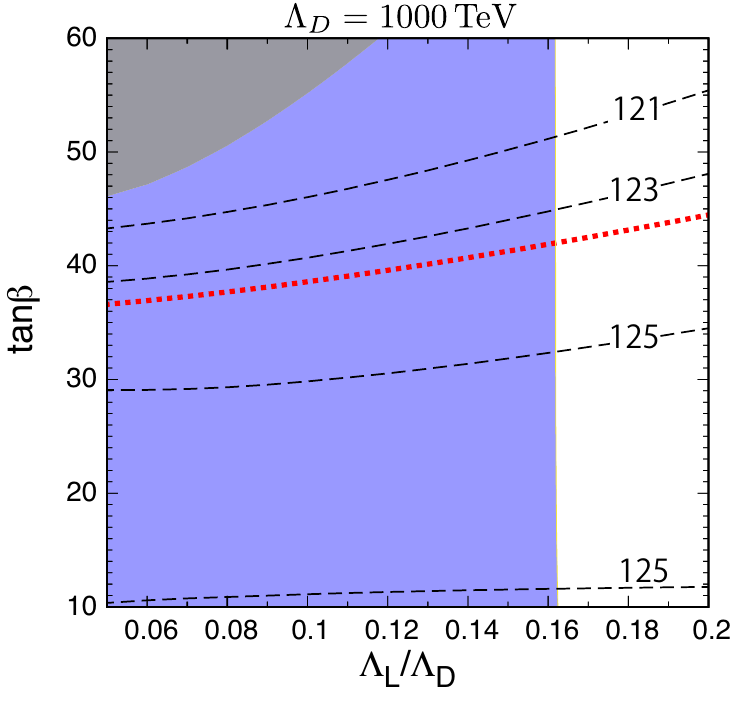}
\caption{
The contours of the Higgs boson mass in units of GeV (black dashed) and 
the SUSY contribution to the muon $g-2$ in the cases of ${\bf 5} + \bar {\bf 5}$ messenger fields. 
In the orange (yellow) regions, the muon $g-2$ is explained at 1$\sigma$ (2$\sigma$) level. 
The regions above red dotted lines are excluded, since the charge breaking minimum of the stau-Higgs potential is problematic.
In the blue shaded regions, the wino, $m_{\chi^{\pm}_1/\chi_1^0}<430$ GeV.
The messenger scale is taken to be $M_L=M_D=1500\,$TeV.
Here, $\alpha_s(m_Z)=0.1185$ and $m_t({\rm pole})=173.34$\,GeV.
}
\label{fig:pre}
\end{center}
\end{figure}

\begin{figure}[!t]
\begin{center}
\includegraphics{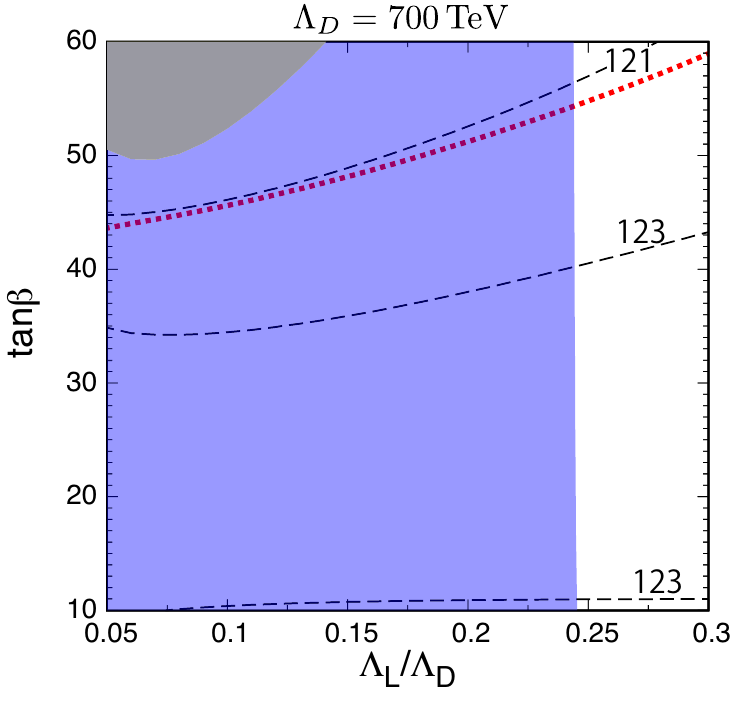}
\caption{
The contours of the Higgs boson mass and the SUSY contribution to the muon $g-2$ in the cases of ${\bf 5} + \bar {\bf 5}$ messenger fields, with a large messenger mass; $M_L=M_D=10^{13}$\,GeV. The other parameters are same as in Fig.~\ref{fig:pre}.
}
\label{fig:pre2}
\end{center}
\end{figure}

In Fig.\,\ref{fig:pre}, we show the contours of the Higgs boson mass and the SUSY contribution to the muon $g-2$. 
Mass spectra of SUSY particles are calculated using {\tt SOFTSUSY 3.7.4}~\cite{softsusy} with appropriate modifications; the Higgs boson mass and the SUSY contribution to the muon $g-2$ are computed using {\tt FeynHiggs 2.13.0}~\cite{feynhiggs,feynhiggs2,feynhiggs3,feynhiggs4,feynhiggs5}.
The messenger scales are taken to be $M_L=M_D=1500\,$TeV.
In the orange (yellow) regions, the muon $g-2$ is explained at 1$\sigma$ (2$\sigma$) level (see Eq.\,(\ref{eq:gm2})). 
The regions above red dotted lines are excluded, since the constraint in Eq.~(\ref{eq:chbreaking}) is not satisfied. 
In the blue shaded regions, the wino, $m_{\chi_1^{\pm}/\chi_1^0}<430$ GeV, which is the excluded region given by the recent LHC wino search~\cite{ATLAS_wino}. By combining the LHC results and the vacuum stability constraint, it can be seen that  there is no region consistent with the muon $g-2$ at 2$\sigma$ level. 
For a larger messenger scale, it is more difficult to explain the muon $g-2$ as shown in Fig.~\ref{fig:pre2}. This is because the sleptons become heavier for the fixed neutralino/chargino mass due to radiative corrections from gaugino loops. In Fig.~\ref{fig:pre2}, the messenger scale is taken as $M_L=M_D=10^{13}$\,GeV. It can be seen that the SUSY contribution to the muon $g-2$ decreases compared to the case of the low messenger scale.
The situation is not improved in the case of  ${\bf 10}$ and $\overline{ {\bf 10}}$ messengers.

The superpotential in the case with {\bf 10}$+$$\overline{\bf 10}$ messenger fields is given by
\begin{eqnarray}
W = (M_Q + k_Q Z) \Psi_Q \Psi_{\bar Q} +(M_U + k_U Z) \Psi_U \Psi_{\bar U} 
+ (M_E + k_E Z) \Psi_E \Psi_{\bar E},
\end{eqnarray}
where $\Psi_Q$, $\Psi_{\bar U}$ and $\Psi_{\bar E}$ are transformed as ({\bf 3}, {\bf 2}, 1/6), ($\bar {\bf 3}$, 1, $-$2/3) and ({\bf 1},{\bf 1},1) under the SM gauge group, respectively.
The soft SUSY breaking masses are generated as
\begin{eqnarray}
M_1 \simeq \frac{g_1^2}{16\pi^2} \frac{1}{5}(\Lambda_Q + 8 \Lambda_U + 6 \Lambda_E), \ 
M_2 \simeq \frac{g_2^2}{16\pi^2} (3 \Lambda_Q), \ 
M_3 \simeq \frac{g_3^2}{16\pi^2} (2 \Lambda_Q +  \Lambda_U), \ 
\end{eqnarray}
\begin{eqnarray}
m_i^2 \simeq
\frac{2}{(16\pi^2)^2} \left[ {C_2(r_3)} g_3^4 (2 \Lambda_Q^2 + \Lambda_U^2)  
+ C_2(r_2) g_2^4 (3 \Lambda_Q^2 ) 
+ \frac{3}{5}Q_Y^2 g_1^4 ( \tilde \Lambda_{10}^2) \right],
\end{eqnarray}
where $\Lambda_Q=k_Q \left<F_Z\right>/M_Q$, $\Lambda_U=k_U \left<F_Z\right>/M_U$, $\Lambda_E=k_E \left<F_Z\right>/M_E$ and $\tilde \Lambda_{10}^2 = (\Lambda_Q^2 + 8\Lambda_U^2 + 6\Lambda_E^2)/5$. In this GMSB model, 
the heavy squarks can be obtained by considering large $\Lambda_U \gg \Lambda_E$; however, $\Psi_{U}$ messenger has a hypercharge larger than that of $\Psi_{\bar D}$ in the previous case. Thus, the sleptons are heavier compared with in the  {\bf 5} + $\bar {\bf 5}$ messenger case for the fixed stop mass and it is not possible to explain the muon $g-2$ anomaly.

\vspace{12pt}
Finally, let us consider the gauge mediation model with the adjoint messengers, which are in  a ${\bf 24}$ representation of $SU(5)$ GUT gauge group. The superpotential is given by
\begin{eqnarray}
W = (M_O + k_O Z) {\rm Tr}(\Sigma_8^2) +(M_T + k_T Z) {\rm Tr}(\Sigma_3^2) + (M_X + k_X Z) X \bar X,
\end{eqnarray}
where $\Sigma_8$, $\Sigma_3$, $X$ and $\bar X$ are transformed as ({\bf 8},{\bf 1},0), ({\bf 1},{\bf 3},0), 
({\bf 3},{\bf 2},$-5/6$) and ({\bf 3},{\bf 2},5/6) under the SM gauge group. 
By integrating out the messenger fields, 
the soft SUSY breaking masses are obtained as
\begin{eqnarray}
M_1 \simeq \frac{g_1^2}{16\pi^2} (5 \Lambda_X), \ 
M_2 \simeq \frac{g_2^2}{16\pi^2} (3 \Lambda_X + 2 \Lambda_3), \ 
M_3 \simeq \frac{g_3^2}{16\pi^2} (2 \Lambda_X + 3 \Lambda_8), \ 
\end{eqnarray}
\begin{eqnarray}
m_i^2 \simeq
\frac{2}{(16\pi^2)^2} \left[ {C_2(r_3)} g_3^4 (3 \Lambda_8^2 + 2\Lambda_X^2)  
+ C_2(r_2) g_2^4 (2 \Lambda_3^2 + 3 \Lambda_X^2 ) 
+ \frac{3}{5}Q_Y^2 g_1^4 (5 \Lambda_X^2) \right],
\end{eqnarray}
where $\Lambda_X = k_X \left<F_Z\right>/M_X$, $\Lambda_3 = k_T \left<F_Z\right>/M_T$, 
$\Lambda_8 = k_O \left<F_Z\right>/M_O$.
In this gauge mediation model, the squarks can be easily made heavier than the sleptons by taking $\Lambda_8 \gg \Lambda_3, \Lambda_X$. This feature is favored for the explanation of the muon $g-2$. However, the model has a serious problem as shown below.

\begin{figure}[!t]
\begin{center}
\includegraphics{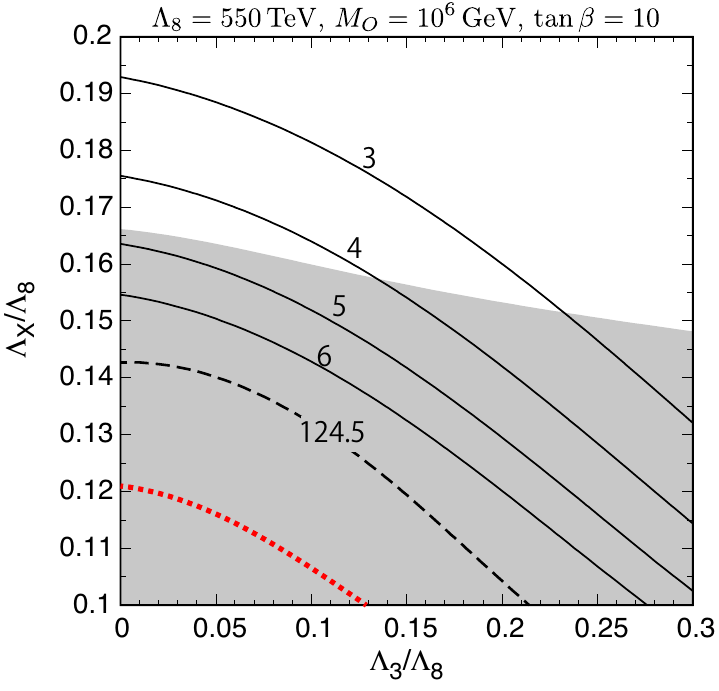}
\includegraphics{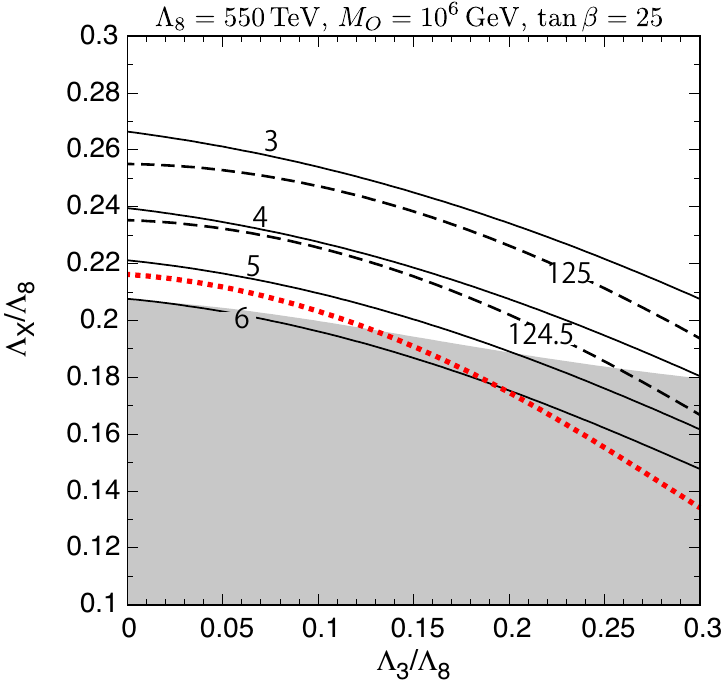}
\caption{
The contours of the SUSY contribution to the muon $g-2$ in units of $10^{-10}$ (black solid) for $M_O=M_T=M_X=10^{6}$\,GeV and $\Lambda_8=550$\,TeV, in the adjoint {\bf 24} model. 
The black dashed lines show the the Higgs boson mass in units of GeV.
In the gray region, the mass of the NLSP stau is smaller than 360\,GeV. 
In the region below the red-dotted line, the life-time of the EWSB minimum is too short and problematic.
Here, $\tan\beta=10$\,(25) on the left (right) panel. 
The other parameters are same as in Fig.~\ref{fig:pre}.
}
\label{fig:pre3}
\end{center}
\end{figure}

\begin{figure}[!t]
\begin{center}
\includegraphics{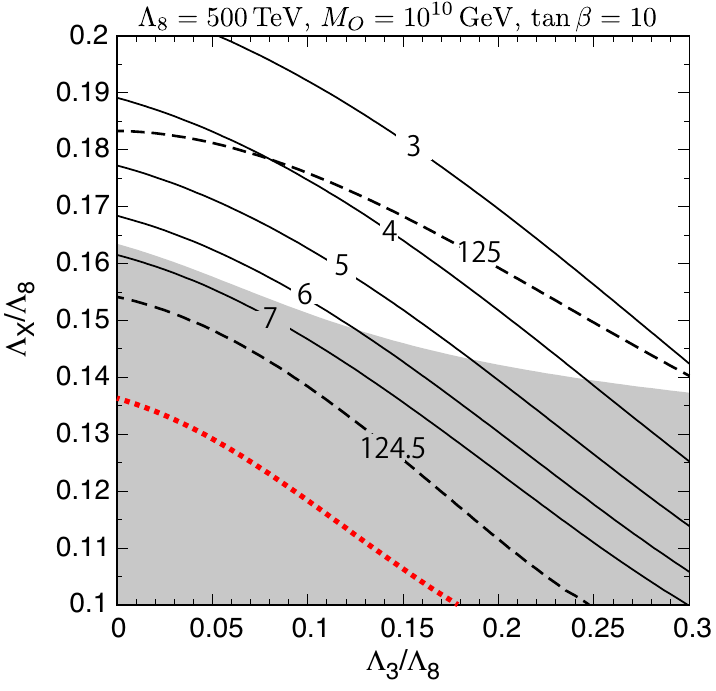}
\includegraphics{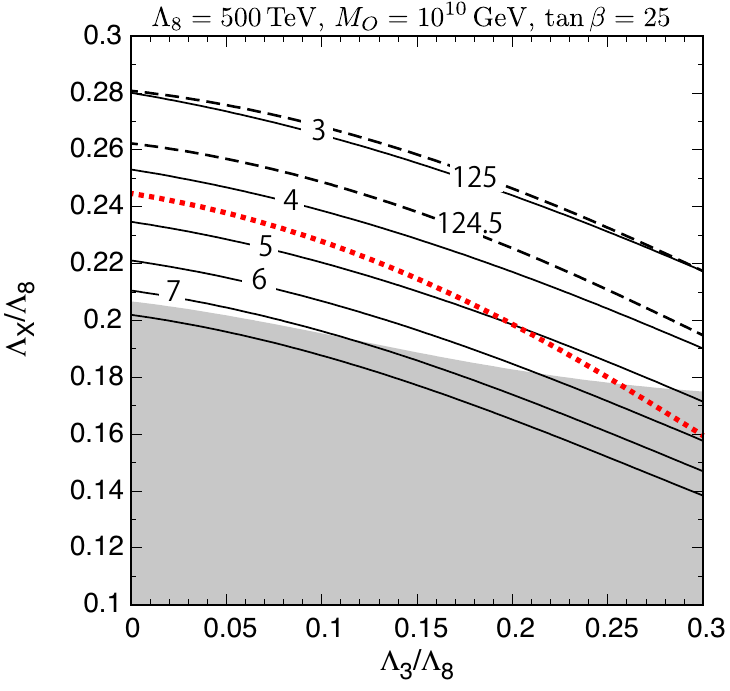}
\caption{
The contours of the SUSY contribution to the muon $g-2$ in units of $10^{-10}$ (black solid) for $M_O=M_T=M_X=10^{10}$\,GeV and $\Lambda_8=500$\,TeV, in the adjoint {\bf 24} model. 
Here, $\tan\beta=10$\,(25) on the left (right) panel. 
The other parameters are same as in Fig.~\ref{fig:pre}.
}
\label{fig:pre4}
\end{center}
\end{figure}

In this GMSB model, only $X$ and $\bar X$ messenger fields contribute to the bino mass and right-handed slepton masses since $\Sigma_8$ and $\Sigma_3$ do not have hypercharges. 
At the messenger scale, the bino and right-handed slepton masses satisfy the relation,
\begin{eqnarray}
\frac{m_{\bar E}^2}{M_1^2} \simeq \frac{1}{6},
\end{eqnarray}
where $m_{\bar E}$ is a mass of a right handed slepton; 
therefore, the right-handed sleptons are lighter than the bino for a small messenger scale. Especially, the stau becomes light ($\lesssim$\,200\,GeV) and the next-to-the lightest SUSY particle (NLSP) in the region where the muon $g-2$ is explained~\cite{Ibe:2012qu}. This stau is long-lived and excluded by the LHC SUSY search~\cite{lhcstau, lhcstau2}. (Here, the wino is slightly heavier than bino at the weak scale even for $k_T=0$.) 
In Figs.~\ref{fig:pre3} and \ref{fig:pre4}, the contours of the Higgs boson mass and the SUSY contributions to the muon $g-2$ are shown 
in the case of {\bf 24} messenger. 
%
In Fig.~\ref{fig:pre3} (\ref{fig:pre4}), the messenger scale is taken as $M_O=M_T=M_X=10^6$\,GeV ($10^{10}$\,GeV) with different values of $\tan\beta$: $\tan\beta=10$ (left panels) and 25 (right panels). 
The gray regions are excluded since the mass of the NLSP stau is smaller than 360\,GeV, which is excluded even if only direct production of the stau is considered~\cite{lhcstau2}.  
Also, the regions below the red dotted lines are excluded due to the vacuum stability constraint in Eq.\,(\ref{eq:chbreaking}). It is shown that the SUSY contribution does not reach $10^{-9}$ due to the LHC constraint and vacuum stability constraint.

\begin{figure}[!t]
\begin{center}
\includegraphics{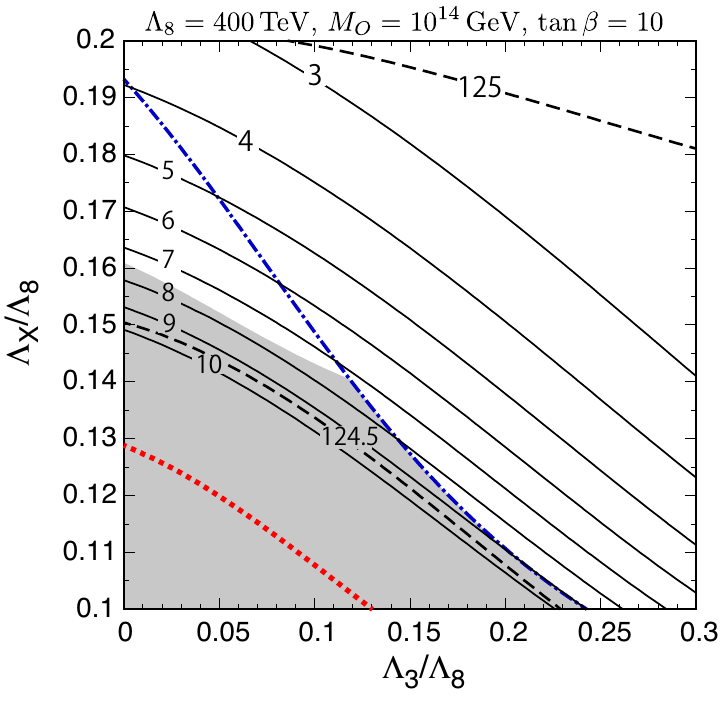}
\includegraphics{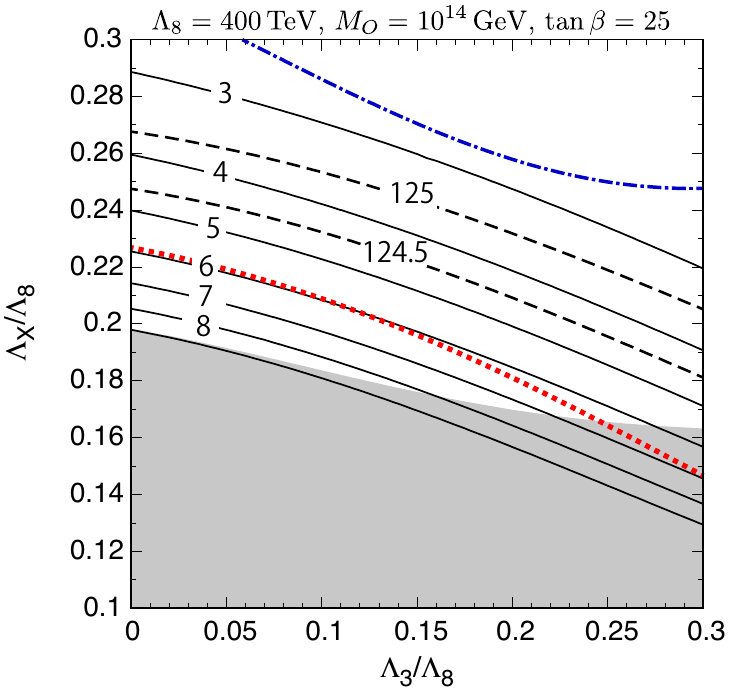}
\caption{
The contours of the SUSY contribution to the muon $g-2$ in units of $10^{-10}$ (black solid) for $M_O=M_T=M_X=10^{14}$\,GeV and $\Lambda_8=400$\,TeV,  in the adjoint {\bf 24} model. 
In the regions above the blue dashed-dotted lines, the lightest neutralino is NLSP. 
Here, $\tan\beta=10$\,(25) on the left (right) panel. 
The other parameters are same as in Fig.~\ref{fig:pre}.
}
\label{fig:pre5}
\end{center}
\end{figure}

For a large messenger scale, there is a region where the stau is heavier than bino. However, the sleptons are not sufficiently light to explain the muon $g-2$ anomaly. This is shown in Fig.~\ref{fig:pre5}. In the figure, the messenger scale is taken to be $M_O=M_T=M_X=10^{14}$\,GeV. On the left (right) panel, $\tan\beta=10$ (25).
In the regions below the blue dashed-dotted lines, the stau is NLSP and long-lived. The gray regions and the regions below the red dotted lines are excluded due to the LHC constraint and the vacuum stability constraint, respectively. 
In this region, the muon $g-2$ anomaly can not be explained.
On the other hand, in the regions above the blue dashed-dotted lines, the stau is heavier than the lightest neutralino. However, the SUSY contribution to the muon $g-2$ is smaller than $10^{-9}$. 

\vspace{12pt}
So far, in the case of the large $\mu$-parameter, the bino contribution to the muon $g-2$ is dominant.
In this case, both of the left-handed and right-handed sleptons should be very light as $\sim$100\,-\,500 GeV. The wino mass should be also small, since the large wino mass makes the left-handed slepton heavy via radiative corrections. This wino is excluded by the recent LHC data. In the case of adjoint {\bf 24} messenger, the constraint from the wino search is avoided; however, the stau is light as $\lesssim$\,200\,GeV and long-lived, which is also excluded.\footnote{ If we introduce soft masses for sleptons generated by gravity mediation, this problem may be solved.}

The above difficulty is solved if there is a mechanism to make the $\mu$-term small. 
This is achieved with the additional Higgs soft mass for $H_u$.
In the small $\mu$ case, the chargino contribution to the muon $g-2$ becomes sufficiently large. Then, the sleptons and wino can be heavier than in the cases of the large $\mu$ parameter. As a result, the muon $g-2$ anomaly is explained consistently with the latest LHC results.

\subsection{A model with additional Higgs soft masses}

Let us consider a simple example which provides the additional Higgs soft masses, $m_{H_u}^2$ and $m_{H_d}^2$. 
We consider the following superpotential~\cite{Ibe:2012qu} (see also \cite{Ibe:2007km})
\begin{eqnarray}
W = m^2 Z + \frac{\kappa}{2} Z X^2 + M_{XY} X Y + \lambda_X X H_u H_d,
\end{eqnarray}
with the minimal K{\" a}hler potential, $K=X^\dag X + Y^\dag Y + Z^\dag Z$. 
After integrating out $X,Y$ fields, we have additional Higgs soft masses:
\begin{eqnarray}
\delta m_{H_{u}}^2 =\delta m_{H_{d}}^2 = \frac{|\lambda_X|^2}{32\pi^2}\frac{ \kappa^2 m^4}{ M_{XY}^2} \left[1  + \frac{\kappa^2 m^4}{6 M_{XY}^4} + \mathcal{O}\left(\frac{\kappa^4 m^8}{M_{XY}^8}\right)\right]. 
\end{eqnarray}
Here, $\mu$, $B_\mu$ and $A$-terms are not generated. 
Note that $Z$ is stabilized at the origin with the Kahler potential
\begin{eqnarray}
\Delta K = -\frac{\kappa^4}{192\pi^2} \frac{|Z|^4}{M_{XY}^2},
\end{eqnarray}
which is induced at the one-loop level; therefore, $\left<Z\right>=0$ is justified. 
Now we can make the $\mu$-parameter small as shown in \cite{Ibe:2012qu}. 
Then, the chargino contribution to the muon $g-2$ can be sufficiently large, which allows the sleptons to be heavier than in the case of the large $\mu$-parameter.

\begin{figure}[!t]
\begin{center}
\includegraphics{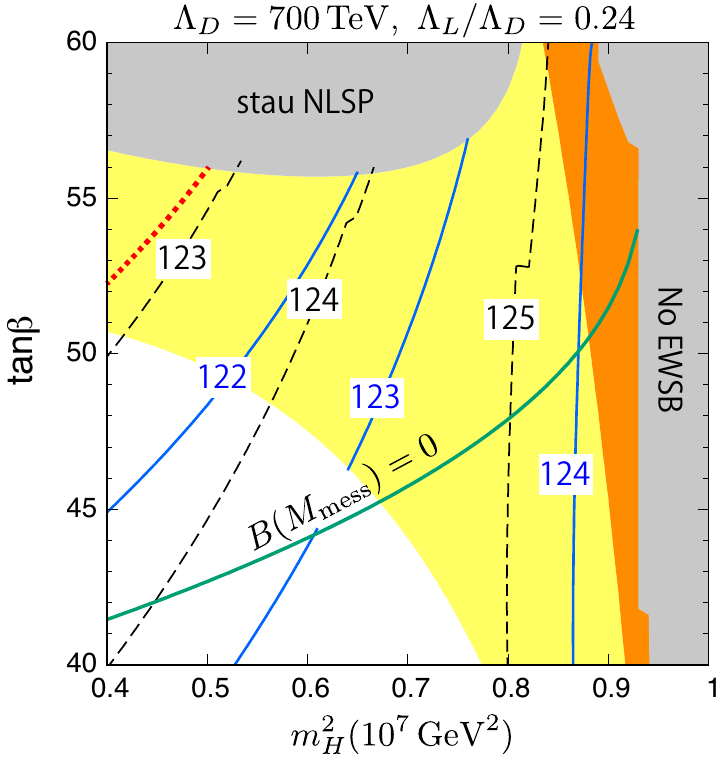}
\includegraphics{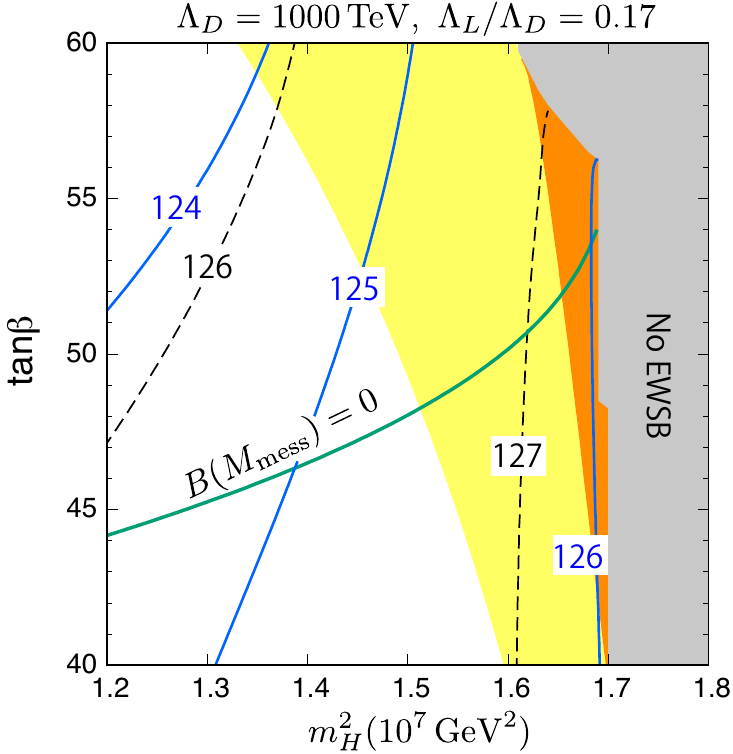}
\caption{
The contours of the Higgs boson mass in units of GeV [black dashed ({\tt FeynHiggs}), blue solid ({\tt SUSYHD})] and the SUSY contribution to the muon $g-2$ with the additional Higgs soft masses.
In the orange (yellow) regions, the muon $g-2$ is explained at 1$\sigma$ (2$\sigma$) level. 
In the region above the red dotted line, the charge breaking minimum of the stau-Higgs potential is problematic.
The messenger scales are taken to be $M_L=M_D=1500\,$TeV.
The other parameters are same as in Fig.~\ref{fig:pre}
}
\label{fig:higgs_gm2}
\end{center}
\end{figure}

\begin{table*}[!t]
\caption{\small Mass spectra in sample points. We take $M_L=M_D$. Here, $\tan\beta$ is determined to satisfy $B (M_D)=0$.
}
\label{tab:sample}
\begin{center}
\begin{tabular}{|c||c|c|c|c|}
\hline
Parameters & Point {\bf I} & Point {\bf II}  & Point {\bf III}  \\
\hline
$M_{D} $ (TeV) & 1500  & 1500  & $1600$  \\
$\Lambda_{D} $ (TeV) & 700  & 1000  & 800  \\
$\Lambda_L/\Lambda_D$  & $0.24$  & $0.19$  & $0.18$ \\
$m_{H}^2 $ ($10^7$\,GeV$^2$) & 0.9  & 1.68  & 1.17  \\
\hline
%
Particles & Mass (GeV) & Mass (GeV)& Mass (GeV)  \\
\hline
$\tilde{g}$ & 4890 & 7070 & 5550   \\
$\tilde{q}$ & 6440 & 9030 & 7270 \\
$\tilde{t}$ & 6420 & 8330 & 6670 \\
$\tilde{\chi}_{1}^\pm$ & 423 & 311  & 226 \\
$\tilde{\chi}_{2}^\pm$ & 559 & 527  & 408 \\
$\tilde \chi_1^0$         & 421 & 305  & 219 \\
$\tilde{\chi}_2^0$        & 520 & 327 & 246  \\
$\tilde \chi_3^0$         & 530 & 526  & 407\\
$\tilde{\chi}_4^0$       & 578 & 763 & 583  \\
$\tilde{e}_{L, R}$       & 680,\,816 & 828,\,1140  & 642,\,914 \\
$\tilde{\tau}_{1,2}$     & 512,\,608&  638,\,866  & 460,\,662 \\
$H^\pm$ & 1470 & 1380 & 1150 \\
$h_{\rm SM\mathchar`-like}$ & 125.5 &  127.2  & 126.4 \\
\hline
$\mu$ (GeV)  & 517  & 314  & 234 \\
$\tan\beta$  & 51.5  & 53.6  & 53.7 \\
$\delta a_\mu (10^{-10})$  & 21.1  & 20.3  & 34.5 \\
\hline
\end{tabular}
\end{center}
\end{table*}

\vspace{12pt}

In Fig.\,\ref{fig:higgs_gm2}, the contours of the Higgs mass and the SUSY contribution to the muon $g-2$ 
on $m_H^2$\,-\,$\tan\beta$ plane are shown, where $m_{H}^2=m_{H_u}^2=m_{H_d}^2$ at the messenger scale. 
Here, a pair of {\bf 5} and $\bar {\bf 5}$ messenger fields is considered.
We take the messenger scales as $M_L=M_D=1500\,$TeV. 
The black dashed (blue solid) lines show computed Higgs boson masses using {\tt FeynHiggs 2.13.0} ({\tt SUSYHD 1.0.2}~\cite{Vega:2015fna}).
In the whole regions, the wino mass is larger than 430\,GeV, although the smaller mass can be allowed for small $\mu$-parameter due to a mixing between the wino and higgsinos.
The black dashed lines show the Higgs boson masses in units of GeV.
In the orange (yellow) regions, the muon $g-2$ anomaly is explained at 1$\sigma$ (2$\sigma$) level. 
The regions above the red dotted lines are excluded, since the charge breaking minimum of the stau-Higgs potential is problematic. Note that on the green solid lines, the Higgs $B$-term vanishes at the messenger scale. Thus, we do not need an additional mechanism to generate the $B$-term.

\vspace{12pt}


Finally, we show mass spectra and the SUSY contribution to the muon $g-2$, $\delta a_\mu$, at some example points in our model in Table~\ref{tab:sample}. In these model points, the Higgs $B$-term vanishes at the messenger scale, i.e., 
the values of $\tan\beta$ are predictions. The Higgs boson masses are evaluated using {\tt FeynHiggs}.
Here, $\mu$-parameter is small as 200\,-\,500\,GeV. On the point {\bf I}, the mass splitting of the wino-like chargino and neutralino is large as 2\,GeV due to their mixing with higgsinos and hence the LHC constraint~\cite{ATLAS_wino} is not applicable.
On the points {\bf II} and {\bf III}, the $\mu$  is smaller than the wino mass. In this case, 
the wino-like chargino/neutralino decays to the higgsino-like neutralino or chargino emitting $Z$, $W$, or the Higgs boson. The LHC constraints on this case~\cite{CMS_chargino_WZ1, CMS_chargino_WZ2} are avoided on the point {\bf II} and {\bf III}.

\section{Conclusion}

We have shown that the muon $g-2$ anomaly can not be explained without an additional Higgs soft mass in the GMSB models where the messenger fields are in ${\bf 5}+\bar{\bf 5}$, ${\bf 10} + \overline{\bf 10}$ or ${\bf 24}$ representation of $SU(5)$.
The difficulty originates from the large $\mu$-term, which is a consequence of the large stop masses required for the Higgs boson of 125 GeV. With the large $\mu$-parameter, the chargino contribution to the muon $g-2$ is suppressed and the sleptons, bino and wino need to be particularly light. Consequently, the required mass splitting between colored and non-colored SUSY particles becomes large. Even if one can achieve the mass splitting, the regions consistent with the muon $g-2$ are excluded by the updated result of the ATLAS wino search. 

It has been shown that the problem is solved if there is an additional Higgs soft mass making the predicted $\mu$-parameter small. In this case, the constraints from the wino search can be easily avoided and the muon $g-2$ anomaly is explained at 1$\sigma$ level.
The fine-tuning to explain the EWSB scale can be taken among the soft SUSY breaking mass parameters.
In this case, the $\mu$-parameter is expected to be as small as $\sim$100\,GeV and consequently 
the light higgsino can be produced at the International Linear Collider experiment.
We also show that there are consistent solutions where the Higgs $B$-term vanishes at the messenger scale, that is, no additional mechanism to generate the $B$-term is required.

\section*{Acknowledgments}
This work is supported by JSPS KAKENHI Grant Numbers 
JP26104009 (T.T.Y), JP26287039 (T.T.Y.), JP16H02176 (T.T.Y),
JP15H05889 (N.Y.) and JP15K21733 (N.Y.);
and by World Premier International Research Center Initiative (WPI Initiative), MEXT, Japan (T.T.Y.).

\end{document}